\documentclass[runningheads]{svmult}
\usepackage{graphicx}  
\usepackage{subeqnar}  
\usepackage{cropmark} 
\usepackage{physprbb}  
\begin{document}
%

%
\thispagestyle{empty}
\begin{flushright}
\large
September 1999
\end{flushright}
\vspace{1.2cm}

\renewcommand{\thefootnote}{\fnsymbol{footnote}}
\setcounter{footnote}{3}
\begin{center}  \begin{Large} \begin{bf}
\hbox to\textwidth{\hss
{Polarized Parton Densities and Processes\footnote{Lecture notes to
appear in `{\em New Trends in HERA Physics 1999}',
Ringberg Castle, Tegernsee, G.\ Grindhammer, B.\ Kniehl, and
G.\ Kramer (eds.) [Springer Lecture Notes]}}\hss}
\end{bf}  \end{Large}

\vspace*{1.8cm}
{\Large M.~Stratmann\footnote{Present address: Institut f\"{u}r Theoretische
Physik, Universit\"{a}t Regensburg, D-93040 Regensburg, Germany}}

\vspace*{8mm}

{Department of Physics, University of Durham, Durham DH1 3LE, England}\\

\vspace*{25mm}

\normalsize
{\large \bf Abstract}
\end{center}
\noindent

The main goals of `spin physics' are recalled, and some 
theoretical and phenomenological aspects of longitudinally polarized 
deep inelastic scattering and other hard processes are reviewed. 
The spin dependent parton densities of protons and photons and 
polarized fragmentation functions are introduced,
and the relevant theoretical framework in next-to-leading order QCD
is briefly summarized. Technical complications typical for spin dependent
calculations beyond the leading order of QCD, like a consistent 
$\gamma_5$ prescription, are sketched,
and some recent results for jet and heavy quark production are discussed.
Special emphasis is put on conceivable measurements at a future 
polarized upgrade of the HERA collider which is currently under consideration.
\setcounter{page}{0}
\renewcommand{\thefootnote}{\arabic{footnote}}
\setcounter{footnote}{0}
\normalsize
\newpage
%
\title*{Polarized Parton Densities and Processes}
\toctitle{Polarized Parton Densities and Processes}
\titlerunning{Polarized Parton Densities and Processes}
\author{Marco Stratmann}
\authorrunning{Marco Stratmann}
\institute{Department of Physics, University of Durham, Durham, DH1 3LE, England}
\maketitle 
\begin{abstract}
The main goals of `spin physics' are recalled, and some 
theoretical and phenomenological aspects of longitudinally polarized 
deep inelastic scattering and other hard processes are reviewed. 
The spin dependent parton densities of protons and photons and 
polarized fragmentation functions are introduced,
and the relevant theoretical framework in next-to-leading order QCD
is briefly summarized. Technical complications typical for spin dependent
calculations beyond the leading order of QCD, like a consistent 
$\gamma_5$ prescription, are sketched,
and some recent results for jet and heavy quark production are discussed.
Special emphasis is put on conceivable measurements at a future 
polarized upgrade of the HERA collider which is currently under consideration.
\end{abstract}

\section{Introduction}
%
One of the most fundamental properties of elementary particles is their spin.
However, the vast majority of past and present experiments at high energy
$e^+e^-$, $ep$, and $pp$ colliders are performed with unpolarized
beams thus neither exploiting the advantages of polarization,
which were demonstrated, e.g., by the SLD experiment at SLAC,
nor revealing any information on the spin dependence of fundamental interactions.
Unlike lepton beams it is an extremely challenging task to
maintain the polarization of protons throughout the acceleration to high
energies, which explains the lack of polarized $ep$ or $pp$ collider 
experiments in the past.
To circumvent this problem, a series of fixed target experiments 
with longitudinally polarized lepton beams scattered off, e.g., proton
targets have been performed at comparatively low energies 
over the past few years \cite{ref:disdata}.

Aiming at polarized deep inelastic scattering (DIS) these experiments 
have been used to extract first information about the 
spin dependent parton densities
\begin{equation}
\label{eq:pdfdef}
\Delta f^H(x,Q^2) \equiv f_+^{H_{+}}(x,Q^2) -  
f_-^{H_{+}}(x,Q^2)\;\;,
\end{equation}
where $f_+^{H_{+}}$ $(f_-^{H_{+}})$ denotes the density of a parton 
$f$ with helicity `+' (`$-$') in a hadron $H$ with helicity `+'. 
It is important to notice that the $\Delta f^{H}$ contain information 
{\em different} from that included in the more familiar 
unpolarized distributions $f^{H}$ [defined by taking the sum on the 
r.h.s.\ of (\ref{eq:pdfdef})], and their measurement is
indispensable for a {\em complete} understanding of the partonic 
structure of hadrons.
However, due to the lack of any experimental information apart from DIS 
and the limited kinematical coverage in $x$ and $Q^2$ of the 
available measurements \cite{ref:disdata}, our knowledge of the $\Delta f$ is 
still rather rudimentary compared to the abundance of results on $f$.

Much experimental progress and, hopefully, exciting new results have 
to be expected in the next couple of years.
Most importantly measurements of, for instance, jet, prompt photon, 
and $W$-boson production rates at the recently completed first 
polarized $pp$ collider RHIC will vastly reduce our ignorance of the $\Delta f$.
Ongoing efforts in the fixed target sector by HERMES \cite{ref:dueren}
and (soon) by COMPASS \cite{ref:compass} to study, in particular, 
semi-inclusive DIS and charm production, respectively, will 
contribute to a more complete picture of polarized parton densities as well.
Here we will mainly focus on the prospects of a conceivable future 
polarized upgrade of the HERA $ep$ collider \cite{ref:herapol}, 
which is currently under scrutiny, and highlight
on some important measurements uniquely possible at an $ep$ collider.

Having pinned down the polarized parton densities (\ref{eq:pdfdef}) one 
can study one of the most fundamental aspects of polarization: 
the question of how the spin $S_z$ of non-pointlike
objects like nucleons is composed of the spin of their constituents, 
the quarks and gluons, and their orbital angular momentum $L_z^{q,g}$. 
The total contribution of quarks and gluons to $S_z$ is determined by the 
first moments of (\ref{eq:pdfdef}), 
$\Delta f(Q^2)\equiv\int_0^1 \Delta f(x,Q^2) dx$, 
and $S_z$ can be written as
\begin{equation}
\label{eq:spinsum}
S_z=\frac{1}{2} = \frac{1}{2} \Delta \Sigma(Q^2)+\Delta g(Q^2)+
                  L_z^q(Q^2)+L_z^g(Q^2)\;\;,
\end{equation}
where $\Delta \Sigma\equiv \sum_q (\Delta q+\Delta\bar{q})$ and $Q$ denotes the
`resolution scale' at which the nucleon is probed. The so far unmeasured 
angular momentum contribution $L_z^{q,g}$ has attracted considerable theoretical
interest recently, and it was suggested \cite{ref:ji} that 
deeply virtual Compton scattering $\gamma^*(Q^2)p\rightarrow \gamma p'$ 
in the limit of vanishing momentum transfer $t=(p-p')^2$ {\em may} 
provide first direct information on $L_z^{q,g}$, however this subject is 
beyond the scope of this talk.

The definition of polarized parton densities (\ref{eq:pdfdef}) also 
holds true for the hadronic content of {\em photons}, $\Delta f^{\gamma}$, 
and can be easily extended to the time-like case, i.e., 
spin dependent fragmentation functions, $\Delta D_f$, as well. 
Both densities have been measured in the unpolarized case,
and their $Q^2$ evolution provides an important test of perturbative QCD.
Needless to stress again that a measurement of $\Delta f^{\gamma}$ and
$\Delta D_f$ is required for a complete understanding of 
space- and time-like distributions.
So far $\Delta f^{\gamma}$ is completely unmeasured, and almost nothing is known
experimentally about spin dependent fragmentation. It is argued below that a
polarized HERA would be also an ideal place to learn more about these densities.

Our contribution is organized as follows: 
First we review the spin dependent proton structure and shall 
give an example of a recent QCD analysis of polarized DIS 
data \cite{ref:grsvupdate}. Then 
the framework is extended to the case of $\Delta f^{\gamma}$ and $\Delta D_f$, 
and theoretical models for these densities are introduced.
Next we turn to polarized processes and briefly sketch the basic technical
framework and complications due to the appearance of $\gamma_5$.
Finally we discuss the main results of two recently finished NLO calculations: 
jet \cite{ref:nlojetpp,ref:nlojetep} and 
heavy flavor production \cite{ref:nlohq}. It should be noted that we have to 
omit several interesting topics such as $L_z$, transverse polarization 
and transversity distributions, single spin processes, etc.
Some recent results and references can be found, e.g., in \cite{ref:dis99werner}.

\section{Polarized Proton Structure and DIS}
%
Longitudinally polarized DIS can be described by introducing a structure function 
$g_1$, in analogy to $F_2$ and $F_L$ in the helicity-averaged case. The
NLO expression for $g_1$ reads (suppressing the obvious $x$ and $Q^2$ dependence)
\begin{equation}
\label{eq:g1}
g_1 =  \frac{1}{2} \sum_{q=u,d,s} e_q^2 \left[
\left(\Delta q + \Delta \bar{q}\right) \otimes 
\left(1+\frac{\alpha_s}{2\pi} \Delta C_q\right) + 
\frac{\alpha_s}{2\pi} \Delta g \otimes \Delta C_g \right]\,\,,
\end{equation} 
where $\Delta C_{q,g}$ are the spin dependent Wilson coefficients, and
the symbol $\otimes$ denotes the usual convolution in $x$ space.
From (\ref{eq:g1}) it is obvious that the available inclusive DIS data 
\cite{ref:disdata} can reveal only information on 
$\Delta q + \Delta \bar{q}$, but neither on $\Delta q$ {\em and} 
$\Delta \bar{q}$ nor on $\Delta g$, which enters (\ref{eq:g1}) only
as an ${\cal{O}}(\alpha_s)$ correction.
Thus all QCD analyses \cite{ref:grsv,ref:qcdanalyses,ref:grsvupdate} have
to impose certain {\em assumptions} about the flavor decomposition in order to 
be able to estimate other hard processes for upcoming experiments like RHIC.
Alternatively one can stick, of course, to a comprehensive analysis of quantities 
accessible in polarized DIS \cite{ref:abfr,ref:smcqcd}.

The $\Delta f$ obey the standard DGLAP $Q^2$ evolution equations -- with 
all unpolarized quantities such as splitting functions replaced by their 
spin dependent counterparts (given in \cite{ref:nlosplit1,ref:nlosplit2}) -- 
which are readily solved analytically in 
Mellin $n$ moment space. A subtlety arises in NLO in the 
non-singlet (NS) sector \cite{ref:weber}. 
The independent NS combinations $q_-=q-\bar{q}$ and $q_+\sim q-q'$ evolve 
in the unpolarized and the polarized case with the same but interchanged kernels, 
i.e., $P_{\pm} = \Delta P_{\mp}$. This simply reflects the fact that in the 
unpolarized case the first moment of $q_-$, the number of valence quarks, 
is conserved with $Q^2$, whereas in the polarized case $\Delta q_+(Q^2)$ 
refers to a conserved NS axial vector current.
The $\Delta f$ are constrained by the unpolarized densities via the
positivity condition
\begin{equation}
\label{eq:pos}
\left| \Delta f(x,Q^2) \right| \le f(x,Q^2)\;,
\end{equation}
which is exploited in most of the QCD analyses. Of course, 
the bound (\ref{eq:pos}) is strictly valid only in LO and is subject to 
NLO corrections \cite{ref:afr} because the $\Delta f$ become unphysical, 
scheme dependent objects in NLO.
However the corrections are not very pronounced, in particular at
large $x$ \cite{ref:afr}, the only region where (\ref{eq:pos}) imposes some
restrictions in practice and hence (\ref{eq:pos}) can be used also in NLO.
%
\begin{figure}[thb]
\begin{center}
\includegraphics[width=.8\textwidth]{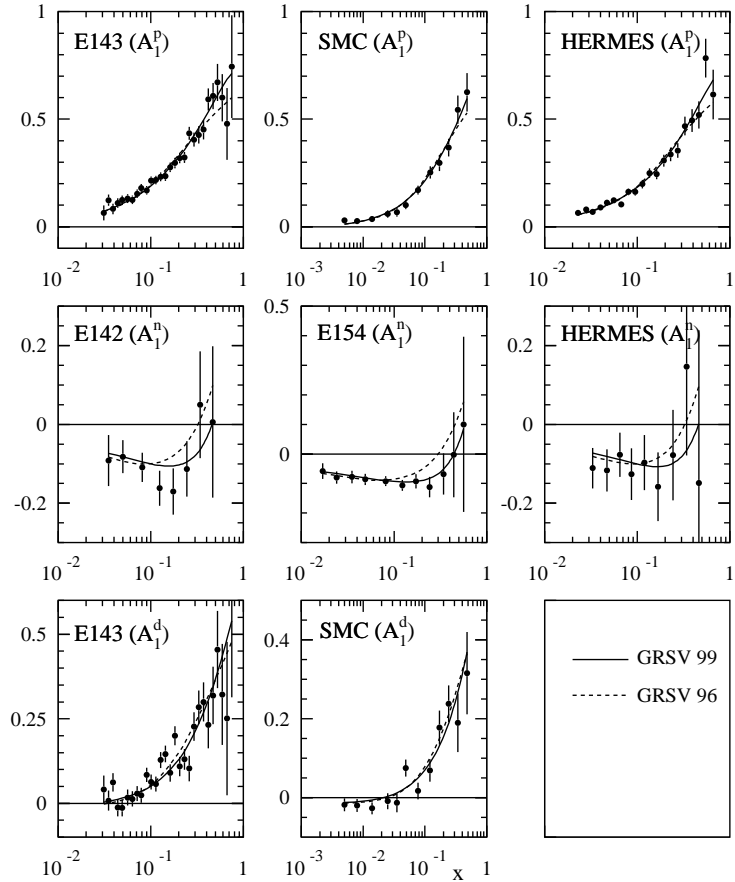}
\end{center}
\vspace*{-3mm}
\caption[]{Comparison of an updated NLO QCD analysis \cite{ref:grsvupdate}
in the GRSV framework \cite{ref:grsv} with available data 
sets \cite{ref:disdata} (the E155 data are not shown, but included in the 
fit). Also shown are the original GRSV results \cite{ref:grsv} based on older 
and fewer data sets.}
\label{fig:fig1}
\end{figure}

Figure~\ref{fig:fig1} shows the result of a recent NLO QCD analysis
\cite{ref:grsvupdate} of all presently available data \cite{ref:disdata}.
The fit is performed directly to the measured spin asymmetry
\begin{equation}
\label{eq:a1}
A_1(x,Q^2)\simeq \frac{g_1(x,Q^2)}{F_2(x,Q^2)/[2x(1+R(x,Q^2))]},
\end{equation}
where $R=F_L/2xF_1$, rather than to the extracted structure function $g_1$ itself.
Eq.~(\ref{eq:a1}) is related to the polarized-to-unpolarized 
cross section ratio $\Delta \sigma/\sigma$, and experimental uncertainties 
like the absolute normalization conveniently drop out. 

As mentioned above, each QCD fit has to rely on several assumptions.
The shown GRSV analysis \cite{ref:grsvupdate} is characterized by the 
choice of a low starting scale for the evolution, $Q_0\approx 0.6\,\mathrm{GeV}$, 
the $\overline{\mathrm{MS}}$ scheme, and a simple but flexible ansatz for the 
polarized densities 
$\Delta f(x,Q_0^2) = N_f x^{\alpha_f} (1-x)^{\beta_f} f(x,Q_0^2)$,
{\em assuming} that $\Delta \bar{q}=\Delta \bar{u}=\Delta \bar{d}$ and
$\Delta s = \Delta \bar{s} = \lambda \Delta \bar{q}$.
For the unpolarized reference distributions $f$ the updated GRV 
densities \cite{ref:grv98} have been used, which also fixes the choice of $Q_0$
(and $\alpha_s(M_z^2)=0.114$). 
The remaining free parameters are determined by the fit after
exploiting constraints for the first moments of the NS combinations
$\Delta q_+$ ($F$ and $D$ values) and by {\em choosing} $\lambda=1$, i.e.,
a $SU(3)_f$ symmetric sea. 
%
\begin{figure}[thb]
\begin{center}
\vspace*{-8mm}
\includegraphics[width=.75\textwidth]{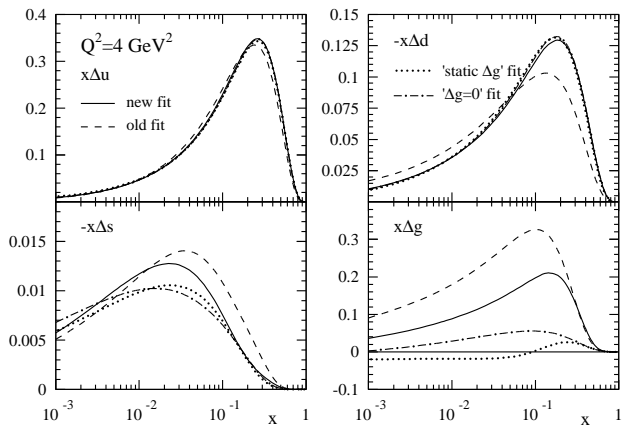}
\end{center}
\vspace*{-3mm}
\caption[]{The polarized NLO $\overline{\mathrm{MS}}$ densities at
$Q^2=4\,\mathrm{GeV}^2$ as obtained in the new \cite{ref:grsvupdate} and 
old \cite{ref:grsv} GRSV analyses. Also shown are the distributions 
obtained in two other fits employing additional constraints on $\Delta g$
(see text).}
\label{fig:fig2}
\end{figure}

The individual parton densities $\Delta f$ resulting from the fit 
in Fig.~\ref{fig:fig1} are shown in Fig.~\ref{fig:fig2}. To demonstrate that, 
in particular, the gluon density is hardly constrained at all by present data, 
two other fits based on additional {\em ad hoc} constraints on $\Delta g$ are 
shown in Fig.~\ref{fig:fig2}.
The `$\Delta g=0$' fit starts from a vanishing gluon input, 
and the `static $\Delta g$' is chosen in such a way that its first moment 
becomes roughly independent of $Q^2$. 
Both gluons give also excellent fits to the available data and do not
affect the results for $u$ and $d$. In fact one can obtain fits 
without changing $\chi^2$ by more than one unit for an even wider range of 
gluon inputs. This uncertainty in $\Delta g$ is compatible with the findings 
of other recent analyses such as 
\cite{ref:smcqcd}. In addition, similarly agreeable fits are obtained, e.g., for 
the choice $\lambda=1/2$ as well as by using an independent $x$ shape for $\Delta s$, 
reflecting the above mentioned uncertainty in the flavor separation.
The range of results for the $\Delta f$ obtained by the various QCD analyses 
\cite{ref:grsvupdate,ref:grsv,ref:qcdanalyses,ref:abfr,ref:smcqcd}
gives a rough measure of the theoretical uncertainties due to different assumptions
used for the fits.
%
\begin{figure}[thb]
\begin{center}
\vspace*{-8mm}
\includegraphics[width=.7\textwidth]{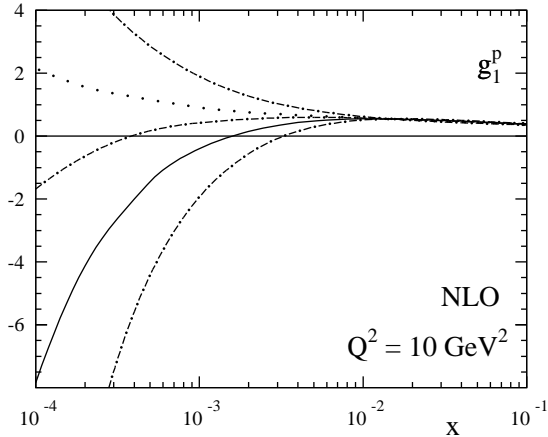}
\end{center}
\vspace*{-4mm}
\caption[]{Predictions for the small $x$ behaviour of $g_1$ 
by extrapolating from the measured region $x>0.01$ to smaller $x$ values for 
different assumptions about $\Delta g$. The solid line is the result 
obtained using the `best fit' $\Delta g$ of \cite{ref:grsvupdate} as shown in
Fig.~\ref{fig:fig2}.}
\label{fig:fig3}
\end{figure}

It is interesting to observe that for the `best fit' gluon in the GRSV 
framework \cite{ref:grsvupdate,ref:grsv}
the spin of the nucleon (\ref{eq:spinsum}) is dominantly carried by 
quarks and gluons at the low bound-state like input scale $Q_0$, and
only during the $Q^2$ evolution a large negative $L_z^g(Q^2)$ is being
built up in order to compensate for the strong rise of 
$\Delta g(Q^2)$, see Fig.~5 in \cite{ref:zeuthen97}.
However, no definite conclusions can be reached yet because 
for the `static $\Delta g$' the situation is completely different, and $S_z$
is entirely of angular momentum origin for {\em all} values of $Q^2$, 
contrary to what is intuitively expected. In addition, direct measurements 
of $L_z^{q,g}$ are completely missing.

Inevitably the large uncertainty in $\Delta g$ implies that the small $x$
behaviour of $g_1$ is completely uncertain and not reliably predictable 
as is illustrated in Fig.~\ref{fig:fig3}. This translates also into a sizeable 
theoretical error for the $x\rightarrow 0$ extrapolation
when calculating first moments of $g_1$, which play an important role in spin
physics since they are related to predictions such as the 
Bjorken sum rule \cite{ref:bjsum}.
The situation is similar to our ignorance of the small $x$ behaviour of $F_2$
in the pre-HERA era and can be resolved only experimentally. Needless to say that
a polarized variant of HERA would be of ultimate help here. In addition, 
the high $Q^2$ region would be accessible for the first time at HERA. 
Here electroweak effects become increasingly important and new structure functions, 
which probe different combinations of parton densities, enter.

\section{Polarized Photon Structure and Fragmentation}
The complete NLO QCD framework for the $Q^2$ evolution of 
$\Delta f^{\gamma}$ and the calculation of the polarized photon
structure function $g_1^{\gamma}$, which would be accessible in $e\gamma$ DIS at a
future polarized linear collider \cite{ref:photon99}, was recently provided in 
\cite{ref:photnlo}.
Unlike the proton densities the $\Delta f^{\gamma}$ 
obey an {\em inhomogeneous} evolution equation schematically given by 
\begin{equation}
\label{eq:photevol}
\frac{d \Delta q_i^{\gamma}}{d \ln Q^2} = \Delta k_i +
\left( \Delta P_i \otimes \Delta q_i^{\gamma} \right) \;\;,
\end{equation}
where $q_i^{\gamma}$ stands for the flavor NS quark combinations 
or the singlet (S) vector 
$\Delta \vec{q}^{\,\gamma}_{S} \equiv {\Delta \Sigma^{\gamma}
\choose \Delta g^{\gamma}}$, and $\Delta k_i$ denotes the photon-to-parton 
splitting functions. Again, solutions of (\ref{eq:photevol}), which can be
decomposed into a `pointlike' (inhomogeneous) and a `hadronic' (homogeneous) part, 
$\Delta q_i^{\gamma} = \Delta q^{\gamma}_{i,PL} + \Delta q^{\gamma}_{i,had}$,
can be given analytically for $n$ moments (cf.~\cite{ref:disgamma}).
It should be noted that perturbative instabilities
for $g_1^{\gamma}$ in the  $\overline{\rm{MS}}$ scheme 
due to the $x\rightarrow 1$ behaviour of the photonic coefficient 
function $\Delta C_{\gamma}$ \cite{ref:photnlo} can be avoided, as in the 
unpolarized case \cite{ref:disgamma}, by absorbing $\Delta C_{\gamma}$ into the 
definition of the quark densities \cite{ref:photnlo} ($\rm{DIS}_{\gamma}$ scheme).
\begin{figure}[thb]
\begin{center}
\vspace*{-8mm}
\includegraphics[width=.75\textwidth]{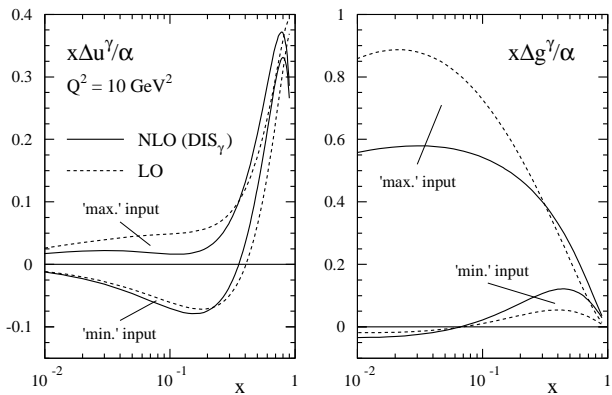}
\end{center}
\vspace*{-6mm}
\caption[]{$x \Delta u^{\gamma}/\alpha$ and $x \Delta g^{\gamma}/\alpha$  
evolved to $Q^2=10\,\mathrm{GeV}^2$ in LO and NLO $(\mathrm{DIS}_{\gamma})$ 
using the two extreme models explained in the text.}
\label{fig:fig4}
\end{figure}

At present the $\Delta f^{\gamma}$ are unmeasured, and one has to fully rely on 
theoretical models. The only guidance is provided by the positivity 
constraint analogous to Eq.~(\ref{eq:pos}). The `current conservation' (CC)
condition \cite{ref:currentcons}, which demands a vanishing first moment 
of $g_1^{\gamma}$ and is automatically fulfilled for the pointlike part
\cite{ref:photnlo}, is not very useful without any data since it can be 
implemented at $x$ values smaller than the one is interested in, say, at $x<0.005$.
To obtain a realistic estimate for the theoretical uncertainties in 
$\Delta f^{\gamma}$ coming from the unknown hadronic input, 
one can consider two very different models \cite{ref:photmodels,ref:photnlo} 
by either saturating the positivity bound (\ref{eq:pos}) at 
$Q_0 \simeq 0.6\,{\mathrm{GeV}}$ (`maximal scenario')
with the phenomenologically successful unpolarized GRV photon densities
\cite{ref:grvphot} or by using a vanishing input (`minimal scenario').
The resulting $\Delta f^{\gamma}$ for both scenarios are shown in 
Fig.~\ref{fig:fig4} and will be applied below to estimate the prospects
of measuring  $\Delta f^{\gamma}$ in photoproduction processes 
at a polarized HERA in the future.

Studies of {\em spin transfer reactions} could provide further invaluable 
insight into the field of spin physics.
A non-vanishing twist-2 spin transfer asymmetry requires the measurement of 
the polarization of one outgoing particle, in addition to having a polarized 
beam or target, and is sensitive to spin dependent {\em fragmentation}.
$\Lambda$ baryons are particularly suited for such studies due to the 
self-analyzing properties of their dominant weak decay, which were 
successfully exploited at LEP \cite{ref:lambdadata} to reconstruct the 
$\Lambda$ spin. In \cite{ref:dsv} a first attempt was made to extract 
the spin dependent $\Lambda$ fragmentation functions, $\Delta D_f^{\Lambda}$,
by analyzing these data \cite{ref:lambdadata}, which, however, turned out to 
be insufficient. Rather different, physically conceivable
scenarios appear to describe the data equally well, and for the `unfavoured' 
sea quark and gluon fragmentation functions one has to fully rely on mere 
assumptions. Clearly, further measurements are required 
to test the models proposed in \cite{ref:dsv}, and, again, HERA can  play
an important role here.
\begin{figure}[thb]
\begin{center}
\vspace*{-17mm}
\includegraphics[width=.75\textwidth]{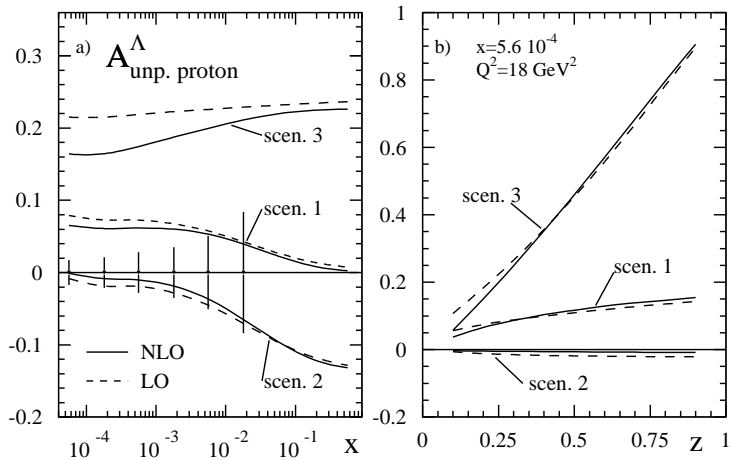}
\end{center}
\vspace*{-15mm}
\caption[]{The semi-inclusive DIS asymmetry $A^{\Lambda}$ for 
unpolarized protons and polarized $\Lambda$'s and leptons 
for the three distinct scenarios of $\Delta D_f^{\Lambda}$ of \cite{ref:dsv}.
In {\bf{a)}} the expected statistical errors for such a measurement at HERA
are shown, assuming a luminosity of $500\,{\mathrm{pb}}^{-1}$, a lepton 
beam polarization of $70\%$, and a $\Lambda$ detection efficiency of 0.1.}
\label{fig:fig5}
\end{figure}

The time-like (TL) $\Delta D_f^{\Lambda}$ are defined in a similar
way as their space-like (SL) counterparts in Eq.(\ref{eq:pdfdef}) via
\begin{equation}
\label{eq:fragdef}
\Delta D_f^{\Lambda} (z,Q^2) \equiv 
D_{f_+}^{\Lambda_+}(z,Q^2) - D_{f_+}^{\Lambda_-}(z,Q^2)\;\;,
\end{equation}
where, e.g., $D_{f_+}^{\Lambda_+}(z,Q^2)$ is the 
probability for finding a $\Lambda$ baryon with positive helicity in a 
parton $f$ with positive helicity at a mass scale $Q$, carrying a fraction 
$z$ of the parent parton's momentum.
The $Q^2$ evolution of (\ref{eq:fragdef}) is similar to the SL case, and
it should be recalled only that the off-diagonal entries in
the singlet evolution matrices $\Delta \hat{P}^{(SL,TL)}$ interchange 
their role when going from the SL to the TL case, see, e.g., 
\cite{ref:furmanski,ref:nlopoltl}.

As a manifestation of the so-called Gribov-Lipatov relation 
\cite{ref:gribov} the SL and TL splitting functions are equal in LO. 
Furthermore they are related by analytic continuation (ACR)
of the SL splitting functions (Drell-Levy-Yan relation \cite{ref:drell}), 
which can be schematically expressed as $(z<1)$
\begin{equation}
\Delta P_{ij}^{(TL)}(z) = z {\cal AC} \Bigg[ \Delta P_{ji}^{(SL)} 
(x=\frac{1}{z}) \Bigg] \; ,
\end{equation}
where the operation ${\cal AC}$ analytically continues any function to 
$x \rightarrow 1/z >1$ and correctly adjusts the color factor 
and the sign \cite{ref:nlopoltl}.
The breakdown of the ACR beyond the LO in the ${\overline{\mathrm{MS}}}$ scheme
can be understood in terms of a corresponding breakdown for the $n=4-2\varepsilon$
dimensional LO splitting functions and can be easily accounted for by a simple
factorization scheme transformation \cite{ref:nlopoltl}.
Alternatively, the ACR breaking can be calculated, of course, graph-by-graph 
\cite{ref:nlopoltl} in the light-cone gauge method \cite{ref:lightcone},
which is of course much more cumbersome.

LO and NLO predictions for the semi-inclusive spin asymmetry $A^{\Lambda}$ 
for the production of polarized $\Lambda$'s in DIS of 
{\em unpolarized protons} off polarized leptons \cite{ref:dsv}
is shown in Fig.~\ref{fig:fig5} for three different conceivable models of
the $\Delta D_f^{\Lambda}$ mentioned above (see \cite{ref:dsv} for details).
Such types of spin measurements, which would help to pin down the 
$\Delta D_f^{\Lambda}$ more precisely, can be performed at HERA 
immediately after the spin rotators in front of H1 and ZEUS 
have been installed even {\em without} having a polarized proton beam.
Similar studies can be done in the photoproduction case where 
an integrated luminosity of only about $100\,\mathrm{pb}^{-1}$ would be 
sufficient \cite{ref:lambdaphot}. Helicity transfer reactions 
can be also examined in $pp$ collisions at RHIC \cite{ref:lambdarhic}.

\section{Polarized Processes}
%
\subsection{Some General Remarks, $\gamma_5$, and All That}
%
To calculate longitudinally polarized cross sections one has to
project onto the two independent helicity configurations of the incoming polarized 
partons (for simplicity we ignore here helicity transfer processes where the
formalism applies in a similar way). This is achieved by using the 
standard relations (see, e.g., \cite{ref:craigie})
\begin{equation}
\label{eq:polgluon}
\epsilon_{\mu}(k,\lambda)\, \epsilon^*_{\nu}(k,\lambda) =
\frac{1}{2} \left[-g_{\mu\nu} + i \lambda \epsilon_{\mu\nu\rho\sigma}
\frac{k^{\rho} p^{\sigma}}{k\cdot p} \right]
\end{equation}
for incoming bosons with momentum $k$ and helicity $\lambda$, and where
$p$ denotes the momentum of the other incoming particle, and
\begin{equation}
\label{eq:polquark}
u(k,h) \bar{u}(k,h) = \frac{1}{2} \not\! k (1-h \gamma_5)
\end{equation}
for incoming massless quarks with momentum $k$ and helicity $h$.
Using (\ref{eq:polgluon}) and (\ref{eq:polquark}) one can calculate the 
cross sections for unpolarized {\em{and}} polarized beams {\em{simultaneously}} 
by taking the sum or the difference of the two helicity dependent 
squared matrix elements
\begin{eqnarray}
\label{eq:unpme}
\mathrm{unpolarized:} \quad \overline{\left| M \right|}^{\: 2} &=&
\frac{1}{2} \left[ \left|M\right|^2(++) + \left|M\right|^2(+-)\right]\\
\label{eq:polme}
\mathrm{polarized:} \quad  \Delta\left|M\right|^2 &=&
\frac{1}{2} \left[ \left|M\right|^2(++) - \left|M\right|^2(+-)\right]
\end{eqnarray}
where $\left|M\right|^2(h_1,h_2)$ denotes the squared matrix element for
any of the contributing subprocesses for definite helicities $h_1$ and $h_2$ 
of the incoming particles.
The possibility to recover well-known unpolarized results `for free' is usually 
regarded as a first important check on the correctness of the 
spin dependent results. 

As usual the presence of IR, UV, and collinear singularities
demands some consistent method to make them manifest.
For this purpose one usually works in the well-established framework of $n$
dimensional regularization (DREG), which immediately leads to complications
in the polarized case since both $\gamma_5$ and the totally antisymmetric 
tensor $\epsilon_{\mu\nu\rho\sigma}$ in 
(\ref{eq:polgluon}) and (\ref{eq:polquark}) are genuine
{\em four} dimensional and have no straightforward continuation to
$n\neq 4$ dimensions\footnote{Sometimes a variant of DREG,
dimensional {\em reduction} \cite{ref:dimred} (DRED), is preferred. Here the 
Dirac algebra is performed in {\em four} rather than $n$ dimensions. 
However, extra counterterms have to be introduced to match the UV sectors of 
DREG and DRED \cite{ref:uvmatch1,ref:uvmatch2}. Once this is done 
DREG and DRED are simply related by a factorization scheme transformation
\cite{ref:uvmatch2,ref:dredtrafo,ref:nlosplit2}.}. 
Since the use of a naive anticommuting $\gamma_5$ in $n$ dimensions
is known to lead to algebraic inconsistencies \cite{ref:cfh},
one usually chooses to work in the HVBM scheme \cite{ref:hvbm}, which was
shown to be internally consistent in $n$ dimensions, and its peculiarities will be
briefly reviewed below. 
Alternatively one can stick to an anticommuting $\gamma_5$ by abandoning the 
cyclicity of trace \cite{ref:koerner}.
In this scheme a `reading point' has to be defined from where {\em all} Dirac
traces of a given process have to be started which can be a quite cumbersome
procedure. Another prescription was suggested to handle traces with one $\gamma_5$ 
\cite{ref:larin} by utilizing 
$\gamma_{\mu}\gamma_5=i/(3!) \epsilon_{\mu\nu\rho\sigma} \gamma^{\nu}
\gamma^{\rho}\gamma^{\sigma}$ and contracting the resulting Levi-Civita tensors in
$n$ dimensions. This avoids $(n-4)$ dimensional scalar productions which show
up in the HVBM scheme but results in more complicated trace calculations.
Needless to say that in the end all consistent prescriptions should give the same
result when used appropriately.

In the HVBM scheme \cite{ref:hvbm} the four dimensional definition for
$\gamma_5$ is maintained, and the $\epsilon$-tensor is regarded as a 
genuinely four dimensional object. In this way the $n$ dimensional space is 
splitted up into a four and a $(n-4)$ dimensional subspace, and
$(n-4)$ dimensional scalar products (`{\em hat momenta}') can show up in 
$\left|M\right|^2(h_1,h_2)$ apart from their usual
$n$ dimensional counterparts (i.e., Mandelstam variables).
For single inclusive jet or heavy quark production, e.g., 
one can choose a convenient frame where all non-vanishing $(n-4)$ 
dimensional scalar products can be expressed by a 
single hat momenta combination $\hat{p}^2$.
These terms deserve special attention when performing the $2\rightarrow 3$ 
phase space integrations since the $(n-4)$ dimensional subspace cannot be 
integrated out trivially as in any unpolarized calculation. However, the 
modified phase space can be conveniently written as 
$\mathrm{dPS}_3 = \mathrm{dPS}_{3,\mathrm{unp}}\times {\cal{I}}(\hat{p}^2)$ such
that it reduces to the well-known `unpolarized' phase space formula
$\mathrm{dPS}_{3,\mathrm{unp}}$ for the vast majority of terms in the 
matrix element which do not depend on $\hat{p}^2$; see
\cite{ref:wvdps3,ref:nlohq} for details. 

The remaining calculation is then standard and proceeds in the same way as 
for any unpolarized cross section with one further crucial exception concerning 
the factorization of mass singularities. 
It was observed \cite{ref:nlosplit2} that the LO polarized splitting 
function in $n=4-2\varepsilon$ dimensions in the HVBM prescription, 
$\Delta P_{qq}^{(0),n}$, is no longer equal to its unpolarized counterpart, 
i.e., it violates helicity conservation, 
$\Delta P_{qq}^{(0),n}(x)-P_{qq}^{(0),n}(x)=4 C_F \varepsilon (1-x)$.
This unwanted property has to be accounted for by an additional factorization 
scheme transformation whenever a pole 
$\sim \Delta P_{qq}^{(0)} 1/\varepsilon$ has to
be subtracted \cite{ref:nlosplit2}. When talking about 
the $\overline{\mathrm{MS}}$ scheme in the polarized case in connection with the 
HVBM prescription, it is always understood that this additional transformation 
is already done.

\subsection{Some Recent Results: Jets, Heavy Quarks}
%
\begin{figure}[thb]
\begin{center}
\vspace*{-11mm}
\includegraphics[width=.69\textwidth]{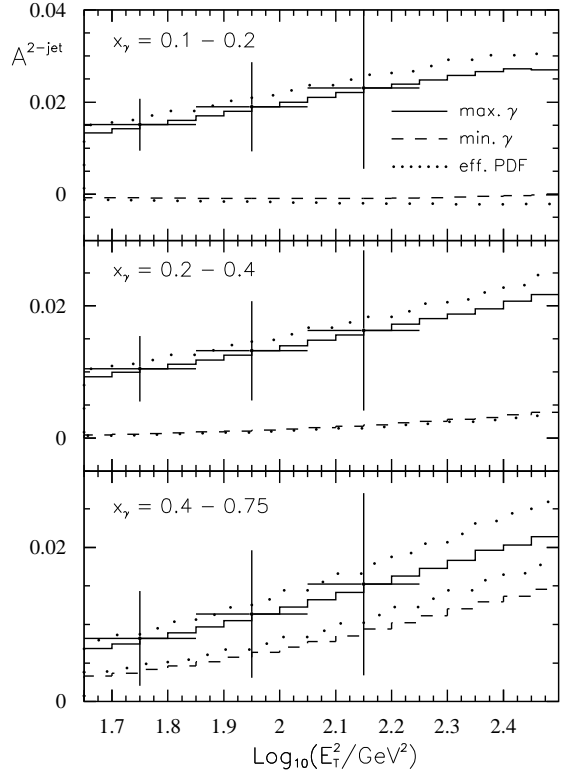}
\end{center}
\vspace*{-7mm}
\caption[]{Predictions for $A^{\mathrm{2-jet}}$ for different
bins in $x_{\gamma}$ using the two scenarios for
$\Delta f^{\gamma}$ as described in the text and the
LO GRSV distributions \cite{ref:grsv} for $\Delta f^p$. 
Also shown are the results using the effective parton density
approximation and the expected statistical errors assuming
a luminosity of $200\,\mathrm{pb}^{-1}$ and $70\%$ beam polarizations.}
\label{fig:fig6}
\end{figure}
Let us finally focus on some recent phenomenological results.
The complete NLO QCD corrections for jet production in
polarized $pp$ \cite{ref:nlojetpp} and $ep$ \cite{ref:nlojetep}
collisions have become available recently in form of MC codes
which allow to study all relevant differential jet distributions.
The photoproduction of jets at a polarized HERA is known to be an excellent 
tool to extract first information on the photonic densities
$\Delta f^{\gamma}$ by experimentally enriching that part of
the cross section that stems from `resolved' photons \cite{ref:svhera}.
In case of single inclusive jet production this can be achieved
by looking into the direction of positive jet rapidities 
(proton direction), and this feature was shown to
be maintained also at NLO \cite{ref:nlojetep}. In addition,
an improved dependence of the cross section on the factorization and 
renormalization scales, $\mu_f$ and $\mu_r$, respectively, 
was found, and the LO jet spin asymmetries in \cite{ref:svhera} receive only
moderate NLO corrections \cite{ref:nlojetep}. 

Similar studies of {\em di}-jet production have the advantage that 
the kinematics of the underlying hard subprocess can be fully reconstructed 
and the momentum fraction $x_{\gamma}$ of the photon can be determined 
on an experimental basis.
In this way it becomes possible to experimentally 
suppress the `direct' photon contribution by introducing some suitable 
cut $x_{\gamma} \leq 0.75$ \cite{ref:jeff},
or by scanning different bins in $x_{\gamma}$.
Very encouraging results were found in \cite{ref:jon}, and it was shown that
the LO QCD parton level calculations nicely agree with `real' jet
production processes including initial and final state QCD radiation
as well as non-perturbative effects such as hadronization, as 
modeled using the spin dependent {\tt SPHINX} MC \cite{ref:sphinx}.
%
\begin{figure}[thb]
\begin{center}
\vspace*{-9mm}
\includegraphics[width=.69\textwidth]{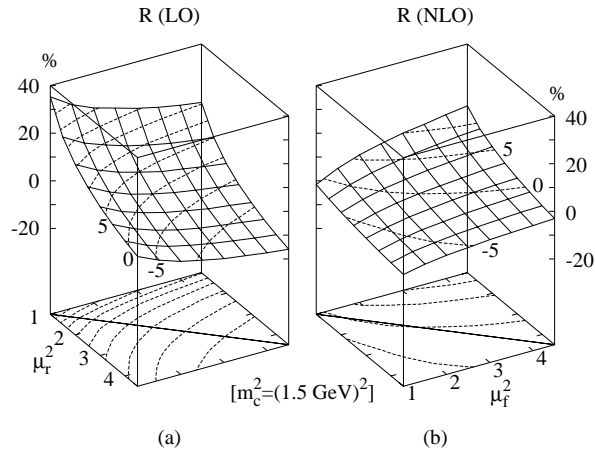}
\end{center}
\vspace*{-7mm}
\caption[]{$R=[\Delta \sigma_{\gamma p}^c(\mu_r^2,\mu_f^2)-
               \Delta \sigma_{\gamma p}^c(\mu_r^2=\mu_f^2=2.5m_c^2)]/
               \Delta \sigma_{\gamma p}^c(\mu_r^2=\mu_f^2=2.5m_c^2)$
in LO {\bf (a)} and NLO {\bf (b)} in percent for $\sqrt{S}=10\,\mathrm{GeV}$.
$\mu_f$ and $\mu_r$ are in units of the charm quark mass $m_c=1.5\,\mathrm{GeV}$.
The contour lines are in steps of $5\%$ and for convenience a line 
corresponding to the usual choice $\mu_f=\mu_r$ is shown at the base of the
plots.}
\label{fig:fig7}
\end{figure}

Figure~\ref{fig:fig6} shows the experimentally relevant di-jet spin asymmetry
$A^{\mathrm{2-jet}}\equiv d\Delta\sigma/d\sigma$ in LO
for three different bins in $x_{\gamma}$ \cite{ref:spin99}, 
using similar cuts as in a 
corresponding unpolarized measurement \cite{ref:effpdfunp}.
$A^{\mathrm{2-jet}}$ in NLO QCD has not been studied yet, but
again only moderate corrections should be expected.
Since it would be a very involved task to unfold the $\Delta f^{\gamma}$ 
from such a measurement of $A^{\mathrm{2-jet}}$
due to the wealth of contributing subprocesses and combinations of 
parton densities, the result of the so-called
`effective parton density approximation' is also shown in Fig.~\ref{fig:fig6}.
This handy but still accurate approximation which is based on an old idea
\cite{ref:effpdf}, allows to straightforwardly extract a specific
`effective' combination of the $\Delta f^{\gamma}$ given by \cite{ref:spin99}
$\Delta f_{\mathrm{eff}}^{\gamma} \equiv 
\sum_q [\Delta q^{\gamma}+ \Delta \bar{q}^{\gamma}]+
\frac{11}{4} \Delta g^{\gamma}$,
once the corresponding combination $\Delta f_{\mathrm{eff}}^p$ for proton 
densities is precisely known from, e.g., jet
production in DIS at a polarized HERA \cite{ref:herapol} or from RHIC.

Finally, the calculation of the NLO QCD corrections to the polarized 
photoproduction of heavy quarks has been finished recently as well 
\cite{ref:nlohq}, and NLO results for the charm contribution $g_1^{\mathrm{charm}}$
to the DIS structure function $g_1$ \cite{ref:nlog1}
and for the hadroproduction of heavy quarks \cite{ref:nlohadro} will become
available very soon. Heavy flavor production is dominated by gluon initiated
fusion processes and hence highly sensitive to the so far poorly
known $\Delta g$. Unfortunately at HERA neither $g_1^{\mathrm{charm}}$
nor the photoproduction of charm give sizeable enough contributions to be
of any use in determining $\Delta g$. In the case of photoproduction of charm
the prospects are much better for the upcoming fixed target experiment 
COMPASS at CERN \cite{ref:compass}.
The NLO corrections in this case appear to be sizeable but well under control, and,
most importantly, the theoretical uncertainties due variations of the
scale $\mu_f$ and $\mu_r$ are greatly reduced when going to the NLO
of QCD \cite{ref:nlohq} as is illustrated in Fig.~\ref{fig:fig7}.

Certainly the next couple of years will produce many new 
experimental results in the field of spin physics.  
In particular first data from the RHIC $pp$ collider, but also results 
from HERMES and COMPASS, will considerably improve our knowledge of the 
spin structure of nucleons. But only a future polarized $ep$ and a 
linear $e^+e^-$ collider can ultimately resolve issues like the small $x$ 
behaviour of $g_1$, the structure of polarized photons, and spin dependent
fragmentation.

\section*{Acknowledgements}
It is a pleasure to thank the organizers for inviting me to
this interesting meeting at such an inspiring location.


\end{document}